\documentclass[prb,onecolumn,aps,amssymb,footinbib,showpacs]{revtex4-1}

\usepackage{eurosym}
\usepackage{amsbsy}
\usepackage{latexsym,epsfig,graphicx}
\usepackage{dcolumn}
\usepackage{graphicx}
\usepackage{subfigure}
\usepackage{comment}
\usepackage{color}
\usepackage{bm}
\usepackage{mathrsfs}
\usepackage{amssymb}
\usepackage{amsthm}
\usepackage{amsfonts}
\usepackage{amsmath}
\usepackage{xspace}
\usepackage{epstopdf}
\usepackage{tabularx}
\usepackage{longtable}
\usepackage[colorlinks=true, letterpaper=true, pdfstartview=FitV, linkcolor=blue, citecolor=blue, urlcolor=blue]{hyperref}
\usepackage[normalem]{ulem}
\usepackage[version=4]{mhchem}

\begin{document}
\title{Tuning the topology of $p$-wave superconductivity in an analytically solvable two-band model}
\author{Haiping Hu$^{1,2}$, Erhai Zhao$^{1}$ and Indubala I. Satija$^{1}$} 
\affiliation{(1) Department of Physics and Astronomy, George Mason University, Fairfax, Virginia 22030, USA}
\affiliation{(2) Department of Physics and Astronomy, University of Pittsburgh, Pittsburgh, Pennsylvania 15260, USA}
\date{\today}

\date{\today}
\begin{abstract}
{
We introduce and solve a two-band model of spinless fermions with $p_x$-wave pairing on a square lattice. The model reduces to the well-known extended Harper-Hofstadter model with half-flux quanta per plaquette and weakly coupled Kitaev chains in two respective limits. We show that its phase diagram contains a topologically nontrivial weak pairing phase as well as a trivial strong pairing phase as the ratio of the pairing amplitude and hopping is tuned. 
Introducing periodic driving to the model, we observe a cascade of Floquet phases with well defined quasienergy gaps and featuring chiral Majorana edge modes at the zero- or $\pi$-gap, or both. 
Dynamical topological invariants are obtained to characterize each phase and to explain the emergence of 
edge modes in the anomalous phase where all the quasienergy bands have zero Chern number.
Analytical solution is achieved by exploiting a generalized mirror symmetry of the model, so that the effective Hamiltonian is decomposed into that of spin-$1/2$ in magnetic field, and the loop unitary operator becomes spin rotations. We further show
the dynamical invariants manifest as the Hopf linking numbers. 
}
\end{abstract}

\maketitle

\section{ Introduction}

Classification of the electronic wave functions according to their topological characteristics
offers fresh insights about band insulators and superconductors described by Bogoliubov-de Gennes (BdG) 
mean-field Hamiltonians \cite{rmp,Qi2011}. 
In two dimensions (2D), integer quantum Hall insulators and spinless $p_x \pm i p_y$  superconductors are two well-known examples of topological phases of 
matter characterized by integer ($\mathbb{Z}$) invariants, the Chern number. 
A topologically nontrivial bulk gives rise to edge modes within the energy gap localized at the sample boundary. More specifically,
the total number of chiral edge states (the net chirality) for a given gap equals to the total Chern number of all the bands below the gap. These edge states are
protected against imperfections, e.g. disorder, as long as certain underlying symmetries of the systems are preserved. 
For example, the $p_x + i p_y$  superconductor has particle-hole symmetry and belongs to class D in the Altland-Zirnbauer symmetry classes
\cite{kitaev,ten-fold}.
Its simplest, single-band model has two phases, the strong pairing phase which is topological trivial with Chern number 0, and 
the topological nontrivial weak pairing phase which has Chern number 1 \cite{PhysRevB.61.10267}.  
The edge excitations in the weak pairing phase are chiral Majorana fermions, 
i.e. their creation operators $\gamma^\dagger_k$ satisfy $\gamma_k^\dagger=\gamma_{-k}$ where $k$ is the momentum along the edge direction.
And at each vortex core, there is localized Majorana zero mode, $\gamma_0^\dagger=\gamma_0$, which obeys non-Abelian statistics
\cite{PhysRevLett.86.268,RevModPhys.80.1083}.

In this paper, we consider a two-band model of spinless $p_x$-wave superconductor in 2D. 
On one hand, our model reduces to the extended Harper-Hofstadter model \cite{harper1955single,hofstadter_energy_1976}
at $\pi$ flux for vanishing superconducting order parameter, $\Delta\rightarrow 0$. 
On the other hand, it reduces to the one-dimensional Kitaev chain \cite{kitaev2001unpaired} 
if the hoppings along the $y$ and diagonal direction are turned off. 
Overall, it incorporates the competition between the band topology, inherited from the familiar integer quantum Hall effect and $p$-wave superconductivity. Our main motivation is to explore its rich phases as the parameters of the model are varied and when the model is time-periodically driven. 
For the static system, we find a weak pairing phase with Chern number 2 that is similar to (but different from) the $p_x + i p_y$ state despite the pairing symmetry is $p_x$-wave, 
and a strong pairing phase that is topologically trivial and can be viewed as stacked Kitaev chains. For the driven (Floquet) system, however,
we discover a myriad of new Floquet phases that feature zero and $\pi$ chiral Majorana edge modes. These phases are characterized not only by the Chern numbers of the bands, but also the three-winding numbers of the unitary evolution operators. 
Thus, this two-band (the number of bands at positive energies) model enriches our understanding of topological superconductivity in 2D in and out of equilibrium, beyond 
the well known one-band model of $p_x + i p_y$ superconductivity.

Despite its apparent complexity relative to the single-band model, we show that this model can be solved analytically for both the static and driven (periodically kicked) cases.
This is achieved by exploiting a generalized ``mirror'' symmetry, which leads to the decomposition of the $4\times 4$ BdG Hamiltonian $H$ into the direct sum of two $2\times 2$ Hamiltonians in the
mirror subspace labeled by $\pm$, $H=H_+\oplus H_-$. In other words, the problem reduces to that of a spin in a momentum-dependent magnetic field described by a Hamiltonian of the form $- {\mathbf{B}}_\pm(\mathbf{k})  \cdot  \boldsymbol{\sigma}$. The decomposition not only greatly simplifies the algebra, but also provides an intuitive picture for all the topological properties, including the Chern number and winding number. It also leads to the definition of mirror-graded Chern numbers.
Furthermore, we show that the nontrivial dynamical topology of the driven model manifests as Hopf links in the $(\mathbf{k},t)$ space.
These appealing analytical and topological properties form the second motivation to introduce our model.
Once Kitaev chain is realized in experiments, 
coupling them together and control the transverse hopping with the help of a synthetic magnetic field will naturally lead to our model.
We stress that our main goal here is {\it not} to propose a model that can be readily realized in experiments. Rather, the toy model is introduced to understand to what extent the topological properties can be, in principle, tuned for a spinless $p$-wave superconductor in 2D while
keeping the problem analytically tractable. 

The paper is organized as follows. In section II we introduce the model, analyze the BdG Hamiltonian in momentum space, and discuss its decomposition in the eigenspace of the mirror symmetry. In section III, we solve the static model to obtain its phase diagram and discuss the topological invariants of each phase (e.g., mirror-graded Chern numbers, the total Chern number) and the corresponding chiral Majorana edge states. We solve the periodically kicked model in Section IV and discuss the Floquet phases, their dynamical invariants, and edge modes in the zero- and $\pi$-gap. An alternative interpretation of the dynamical invariant is presented from the perspective of Hopf links. Section V discusses the implications of our results.
 
\section{ Model}

Consider spinless fermions on a square lattice in two dimensions described by the following tight-binding Hamiltonian,
\begin{eqnarray}
H_s   =  - \sum_{\bm r}  \left [ J_x c_{\bm r+\hat{x}}^{\dag} c_{\bm r} + J_y  e^{i2 n\pi \phi }c_{\bm r+\hat{y}}^{\dag} c_{\bm r}
+J_d  e^{i (n+\frac{1}{2})2 \pi \phi }c_{\bm r+\hat{x}+\hat{y}}^{\dag}c_{\bm r}+J_d e^{i(n-\frac{1}{2})2\pi\phi}c_{\bm r-\hat{x}+\hat{y}}^{\dag} c_{\bm r} 
+\Delta c_{\bm r+\hat{x}}^{\dag} c_{\bm r}^{\dag}  +h.c.\right ] .
\label{Hamil}
\end{eqnarray}
Here $\bm r=n\hat{x}+m\hat{y}$ labels the lattice sites with $n,m$ being integers, $c^{\dag}_{\bm r}$ creates a fermion at site $\bm r$, and we have set the lattice spacing to be unity.  The parameters $J_x$, $J_y$ and $J_d$ denote the hopping along the $x$, the $y$, and the lattice diagonal direction  respectively. There is a magnetic field along the $z$ direction that gives rise to a magnetic flux $\phi$ per unit cell.  In cold atom set ups using optical lattices,  this can be achieved using artificial magnetic field \cite{gauge-rev}. In the Landau gauge, the vector potential $A_x=0$ and $A_y = n \phi h/e$, where $\phi$ is the magnetic flux per plaquette of the square lattice measured in unit of flux quantum $h/e$. We will focus on $\phi=1/2$, i.e., the case of $\pi$-flux. The unit cell then consists of two adjacent plaquette aligned along the $x$ direction. The magnetic Brillouin zone is $k_x\in [0,\pi]$ and $k_y\in[0,2\pi]$. Finally, $\Delta$ is the $p_x$-wave Copper pairing amplitude. We do not determine $\Delta$ self-consistently. Rather we treat it as a tunable parameter controlled by, for example, proximity effect with a nearby $p_x$-wave superconductor. For fixed $m$ (a row of the lattice), the pairing bears the same form as the Kitaev chain  \cite{kitaev2001unpaired}. In fact, the Hamiltonian Eq. \eqref{Hamil} can be viewed as  Kitaev chains extending in the $x$ direction and coupled by $J_y$ and $J_d$ hopping. We have assumed the chemical potential $\mu=0$. And we will measure energy in units of $J_x$.

In the limit of $\Delta=0$, the model reduces to the Harper-Hofstadter model at flux $\pi$ with diagonal hopping $J_d$ \cite{PhysRevB.28.4272,PhysRevB.42.8282}, which induces a gap at energy $E=0$. The two-band model exhibits quantum Hall effect. For small finite $\Delta$, we will show below that the system is a topological superconductor similar to the $p_x+ip_y$ chiral superconductor. For large $\Delta$ the system is topologically trivial. Thus model Eq. \eqref{Hamil} describes the competition between these two phases separated by a topological transition. We will first focus on the static case in section III, where $J_y$ etc. are constant parameters. Later in Section IV, we will consider $J_y$ as function of time $t$, for example,
\[
J_y\rightarrow J_y(t)=J_yT\sum _\ell \delta( t-\ell T),
\]
while all other parameters are held constant. In other words, the system is periodically kicked with time period $T$ \cite{PhysRevLett.112.026805}. This is an example of Floquet, or periodically driven, systems for which $H(t)=H(t+T)$ \cite{bukov2015universal,RevModPhys.89.011004}.

The Hamiltonian appears simpler in momentum space. 
Since the $H_s$ is cyclic in $y$, Fourier transform along the $y$ direction gives the following effective 1D Hamiltonian (with energy in units of $J_x$) for given quasimomentum $k_y$,
\begin{eqnarray}
H({k_y})=-\sum_n \left[J_n c_{n+1,k_y}^{\dag} c_{n,k_y}+ \Delta c_{n+1,k_y}^{\dag}c^{\dag}_{n,-k_y}+h.c.\right] + \sum_n \mu_n  ( c_{n,k_y}^{\dag}c_{n,k_y} - {1}/{2}),
 \label{hop}
\end{eqnarray}
where we have introduced 
\[
J_n   =   1+ 2\alpha \cos \left [ 2\pi \phi (n+{1}/{2}) + k_y \right ]
\]
and
\[
 \mu_n  =  2 \lambda  \cos ( 2 \pi \phi n+k_y).
\]
Here the ratio $\lambda= {J_y}/{J_x} $ is the $x-y$ hopping anisotropy and $\alpha = {J_d}/{J_x}$. For $\pi$-flux, label the two sites within the unit cell as $A$ and $B$. More explicitly,
even $n$ corresponds to the $A$ sites for which
\begin{eqnarray*}
 \mu^A  =  2\lambda \cos k_y,\,\
 J^A  =  1-2\alpha\sin k_y.
 \end{eqnarray*} 
And for $n$ odd, or $B$ sites,
 \begin{eqnarray*}
 \mu^B  =  -2\lambda \cos k_y,\,\
 J^B  =  1+2\alpha\sin k_y.
 \end{eqnarray*} 
We then perform another Fourier transform along the $x$ axis,
\begin{eqnarray*}
c_{j,A}^{\dag} & = & \frac{1}{\sqrt{N}}\sum_{k_x} e^{i k_x 2 j} c_{A,k_x}^{\dag}\\
c_{j,B}^{\dag} & = & \frac{1}{\sqrt{N}}\sum_{k_x} e^{i k_x (2 j+1)} c_{B,k_x}^{\dag},
\end{eqnarray*}
where $j$ labels the unit cell, and $N$ is the total number of unit cells, and $k_x\in[0,\pi]$. Introduce  
Nambu spinor $\Psi_{\bm k}=(c_{A,\bm k},c_{B,\bm k},c^{\dag}_{A,-\bm k},c^{\dag}_{B,-\bm k})^T$,
then the original Hamiltonian becomes
\[
H_s=\frac{1}{2}\sum_{\bm k}\Psi_{\bm k}^{\dag}H_{BdG}(\bm k)\Psi_{\bm k}.
\]
The  Bogoliubov-de Gennes (BdG) Hamiltonian can be expressed in compact form
\begin{eqnarray}
H_{BdG}(\bm k)=-2\lambda   \cos k_y\sigma_z\tau_z-2\cos k_x \sigma_x\tau_z+4\alpha\sin k_x\sin k_y\sigma_y+2\Delta\sin k_x\sigma_x\tau_y.\label{indubdg}
\end{eqnarray}
Here the Pauli matrices $\tau_i$ ($\sigma_j$) are defined within the particle-hole (AB sublattice) space. Products such as $\sigma_x\tau_y$ are understood as
direct product $\sigma_x\otimes\tau_y$ where $\otimes$ is omitted for brevity. A main appeal of our model $H_s$ is that in $\mathbf{k}$ space, it has a nice form
$H_{BdG}(\bm k)$, which facilitates analytical analysis. It is clear that our main motivation here is 
to construct a simple model that contains the interplay of band topology
and $p$-wave superconductivity and can reduce to well-known models in certain limits.

The BdG Hamiltonian is invariant under the transformation $\mathcal{O}=\sigma_y\tau_x$ which switches 
particle to  hole and simultaneously sublattice A to B,
\[
[H_{BdG}(\bm k), \mathcal{O}]=0.
\]
We will refer to $\mathcal{O}$ as a generalized ``mirror'' symmetry \cite{PhysRevLett.111.056403}, or mirror symmetry for short.
This symmetry will play an important role in the topological characterization below.
Let $V$ be the unitary transformation that diagonalizes $\mathcal{O}$, $V^{\dag}\mathcal{O}V=-\tau_z$.
In the eigenbasis of $\mathcal{O}$, the Hamiltonian 
is decomposed into the direct sum \cite{Hu_2019},
\[
V^{\dag}H_{BdG}(\bm k)V=H_{-}(\bm k)\oplus H_{+}(\bm k)
\]
where
\begin{eqnarray}
H_{\pm}(\bm k)  =  2(\pm \lambda  \cos k_y   -\Delta\sin k_x) \sigma_z 
 +   2 \cos k_x\sigma_x \pm 4\alpha\sin k_x\sin k_y\sigma_y
\label{H12}
\end{eqnarray}
are $2\times 2$ matrices, i.e., $\bm{k}$-dependent spin 1/2 Hamiltonians. Here the subscript $\pm$
labels the subspace where $\mathcal{O}$ is diagonalized and has eigenvalue $\pm 1$ respectively. 
The systematic simplification from $H_s$ to $H_{\pm}$ enables us to
solve for its eigenspectra  and obtain its topological invariants. 

\section{ Phase Diagram of the Static Model }

To find the spectra  of the static BdG Hamiltonian, we can express it in a $2\times 2$ block form,
\begin{equation}
 H_{BdG}= 
\left(
\begin{array}{cc}
 D + F & -i G \\
 i G & -D + F 
 \end{array}
 \right)
 \end{equation}
 where
 \begin{eqnarray*}
 D   & =  &  - 2 \lambda \cos k_y {\sigma_z} - 2 \cos k_x {\sigma_x},\\
 F  & =  & 4 \alpha \sin k_x \sin k_y {\sigma_y},\\
 G  & = &  2  \Delta \sin k_x { \sigma_x}.
 \end{eqnarray*}
 Its square takes a very simple form,
 \begin{eqnarray}
 H^2_{BdG} & = &  4 \left [ a + b \left(
\begin{array}{cc}
 0 & {\sigma_y} \\
 {\sigma_y}  & 0 
 \end{array}
 \right) \right ]
 \end{eqnarray}
 where
 \begin{eqnarray*}
  a  & =  & \cos^2 k_x  + \lambda^2 \cos^2 k_y+ 4 \alpha^2 \sin^2 k_x \sin^2 k_y + \Delta^2 \sin^2 k_x, \\
   b & =  & 2 \lambda \Delta \cos k_y \sin k_x.
 \end{eqnarray*}
 Consequently, the eigenvalues $E$ of $H_{BdG}$ are simply  $\pm  E_\pm$ with $E_\pm=2 \sqrt{a \pm b}$ or explicitly, 
 \begin{equation}
 E _\pm  =  2 \sqrt{ \cos^2 k_x+ ( \lambda \cos k_y \mp \Delta \sin k_x)^2+ 4 \alpha^2 \sin^2 k_x \sin^2 k_y }.
 \label{Estatic}
 \end{equation}
Alternatively, we can start directly from the spin 1/2 Hamiltonians $H_{\pm}(\bm k)$ in Eq. \eqref{H12} and arrive at Eq. \eqref{Estatic} immediately.
According to Eq. \eqref{Estatic}, at positive energies there are two overlapping bands $E_\pm(\mathbf{k})$, corresponding to $H_{\pm}$ respectively. The other bands at
negative energies are just the particle-hole reflection of those at $E>0$. The positive and negative bands are separated by a gap as shown in
Fig. 1(a).

The bulk gap closes when $\Delta=\lambda$. The gap closing occurs at two $\mathbf{k}$ points, $\bm k_1=(\pi/2,0)$ and $\bm k_2=(\pi/2,\pi)$. This marks a quantum
phase transition point that separates two gapped phases, a weak pairing (WP) phase at $\Delta<\lambda$ and strong pairing (SP) phase for $\Delta>\lambda$. These two phases
are topologically distinct. Even though the two bands are overlapping in energy, thanks to the $\mathcal{O}$ symmetry and the decomposition Eq. \eqref{H12},
we can define the Chern number $\mathcal{C}_{\pm}$ in each mirror subspace, e.g., for the lower band $-E_\pm(\mathbf{k})$ of  $H_{\pm}$ 
respectively \cite{PhysRevB.90.165114}.
The total Chern number is then 
\[
\mathcal{C}=\mathcal{C}_{+}+\mathcal{C}_{-}.
\]
We find $\mathcal{C}=2$ for weak pairing phase, while $\mathcal{C}=0$ for the strong pairing phase. 
Define the mirror-graded Chern number,
\[
\mathcal{C}_m=\frac{\mathcal{C}_+-\mathcal{C}_-}{2}.
\]
For the weak pairing phase, we find $C_{+}=C_{-}=1$, while for the strong pairing phase, $C_{+}=C_{-}=0$. In both cases, $C_m=0$.
The transition from the weak pairing phase to the strong pairing phase is a topological phase transition where the total Chern number changes by $-2$. 
This is summarized in Table. \ref{Table2}  below. The situation here differs from another model we studied earlier, where the strong pairing phase has 
$\mathcal{C}=2$ and $\mathcal{C}_m=1$ \cite{Hu_2019}. A nonzero value of $C_m$ means the presence of counter-propagating (sometimes
referred to as helical) Majorana edge modes protected by mirror symmetry. 


\begin{table}[h!]
\centering
\newcommand\T{\rule{0pt}{2.5ex}}
\newcommand\B{\rule[-1.7ex]{0pt}{0pt}}
\centering
\begin{tabular}{lcccccc}
\hline\hline 
\;\quad & $\mathcal{C}_{+}$ & $\mathcal{C}_{-}$ & $\mathcal{C}$  & Edge modes \T\\[3pt]
\hline
$\Delta<\alpha$ (WP) & 1 & 1 & $2$ & Chiral Majorana \T\\
$\Delta>\alpha$ (SP) & 0 & 0 & $0$  & Edge dependent  \T\\
\hline\hline
\end{tabular}
\caption{\label{Table2}
{Summary of the topological invariants, defined in the main text, of the static model in the weak pairing (WP) phase and strong pairing (SP) phase. Also shown are the characterization of their edge modes.}}
\end{table}

\begin{figure}[h]
\centering
\includegraphics[width=5 in]{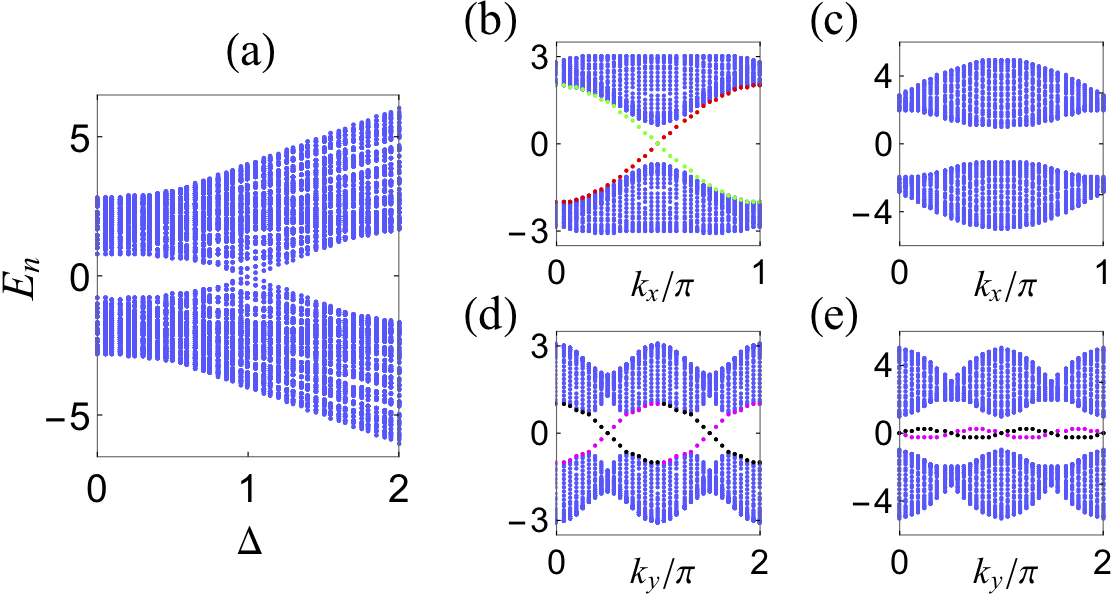}
\caption{ (a) The energy spectrum of the static model Eq. \eqref{Hamil} as $\Delta$ is varied. The topological transition between the weak pairing phase and strong pairing phase occurs at $\Delta=1$ where the gap closes. (b)(c) Energy spectra of the static model in stripe geometry with $y$-open boundary. The red/green dots label the boundary states at the lower/upper boundaries. (d)(e) Energy spectra of the static model in stripe geometry with $x$-open boundary. The magenta/black dots label boundary states at the left/right boundaries. The blue dots label the bulk bands. For (b)(d), $\Delta=0.5$; for (c)(e), $\Delta=1.5$. Other parameters are $\alpha=0.2$, $\lambda=1$. Energy is in units of $J_x$.}
\label{fig1}
\end{figure}

The distinction between the WP and SP phase is reflected in their edge states as illustrated in Fig. \ref{fig1}.
The weak pairing phase is a topological superconductor characterized by $\mathcal{C}_{\pm}=1$.  It is analogous to the $p_x+ip_y$ state, albeit with $\mathcal{C}=2$.
According to the bulk-edge correspondence, there are two chiral Majorana edge modes on the system boundary, see 
Fig. \ref{fig1}(b) and (d). Note that in Fig. \ref{fig1}(b), each edge mode (in color red or green) is two-fold degenerate, in consistent with $\mathcal{C}=2$.
In contrast, the strong pairing phase is topologically trivial. It can still have edge states, but their existence depend on the details, e.g. the orientation of edge, and therefore is not topologically protected.
More specifically, edge states are absent at the upper and lower edges as shown in Fig. \ref{fig1} (c), but they appear 
at the left and right edges as shown in Fig. \ref{fig1} (e).

The edge-dependent boundary modes in the strong pairing phase can be further understood by viewing the 2D lattice system as coupled Kitaev chains. In the limit of $J_y=J_d=0$, each chain along the $x$ direction is a $p_x$-wave superconductor described by Kitaev's model with $\mu=0$, which hosts Majorana zero modes at the chain boundary, say at $x=0$ and $x=L$.
Small but finite  $J_y$ and $J_d$ couple these zero modes to lift their degeneracy. As a result, 
for an $x$-open boundary, the in-gap edge modes acquire dispersion in the $y$ direction but
remain well separated from the bulk band, in agreement with the numerical results in Fig.  \ref{fig1} (e).
Note that in this case there is no net current flow along the edge.
For $y$-open boundaries, each chain is fully gapped and weak inter-chain coupling do not introduce
any in-gap modes. 
Thus, the strong pairing phase as stacked Kitaev chains is a weak
topological phase, but not a genuine 2D topological superconductor.
Its edge-dependent boundary modes are reminiscent of the 1D topology.
For the weak pairing phase at $\Delta<\lambda$, one no longer can treat the interchain coupling as small, and the system is intrinsically two-dimensional.

 \section{ The Periodically Driven Model}
 
Now we generalize the static model  Eq. \eqref{Hamil} to see if periodic driving 
\cite{PhysRevB.79.081406,lindner_floquet_2011,
kitagawa_topological_2010,lindner_topological_2011,kitagawa_transport_2011,jiang_majorana_2011,
gu_floquet_2011,rudner_anomalous_2012}
can induce new topological phases that differ from the weak and strong pairing phase above.
To this end, we make the replacement
\[
J_y\rightarrow J_y(t)=J_yT\sum _\ell \delta( t-\ell T) 
\]
in $H_s$, such that $H_s(t)=H_s(t+T)$. In this periodically kicked model, $J_x$, $J_d$, and $\Delta$ are all held constant, while the hopping $J_y$ is turned on when time $t$ is multiples of the 
kicking period $T$, $t=\ell T$ with $\ell$ an integer. At each time instant, the system is described by $H_{BdG}(\mathbf{k},t)$,
or equivalently its decomposition $H_{\pm}(\mathbf{k},t)$,  obtained from Eq. \eqref{indubdg} by replacement
\[
\lambda \rightarrow \lambda (t)=\lambda T\sum _\ell \delta( t-\ell T) .
\]
Define the time evolution operator as usual,
\[
 U(\mathbf{k},t)=\mathcal{T}e^{-i \int^t_0 H_{BdG}(\mathbf{k},t') dt'}, 
\]
where $\mathcal{T}$ denotes time ordering. In particular, $U(\mathbf{k},T)$ is known as the Floquet operator. Its eigenvalues $e^{i \omega_n(\mathbf{k}) T}$ 
yield the quasienergy band structure $\{\omega_n(\mathbf{k})\}$, where $n$ is the band index and the quasienergy $\omega_n$ is defined modulo $2\pi/T$. Due
 to the $\mathcal{O}$ symmetry, to find the eigenvalues of $U(\mathbf{k},T)$, it 
is sufficient to consider the Floquet operator in the $\pm$ subspace of $\mathcal{O}$, 
\[
 U_\pm(\mathbf{k},T)=\mathcal{T}e^{-i \int^T_0 H_{\pm}(\mathbf{k},t') dt'}=e^{ \mp 2 i T \lambda \cos k_y \sigma_z} e ^{-i T (-2\Delta\sin k_x \sigma_z + 2 \cos k_x\sigma_x\pm 4\alpha\sin k_x\sin k_y\sigma_y) }.
\]
The $2\times 2$ unitary matrix consists of two successive spin rotations, $ U_{\pm}=e^{-i \chi_2\sigma_z}e^{-i\chi_1 (\hat{n}\cdot\boldsymbol{\sigma})}$ where
\begin{eqnarray}
\chi_1  & =  &  T \sqrt{ 4 \cos^2 k_x + 16 \alpha^2 \sin^2 k_x \sin^2 k_y + 4 \Delta^2 \sin^2 k_x},   \nonumber\\
 \chi_2 & =  & \pm 2 T  \lambda \cos k_y,   \nonumber \\
n_x  & = &  { 2 T \cos k_x}/{\chi_1},\\
 n_y &= &\pm   { 4 T \alpha \sin k_x \sin k_y}/{\chi_1},\\
 n_z  & = &  {2 T \Delta \sin k_x}/{\chi_1}.
\end{eqnarray}
Here we have suppressed the subscript $\pm$ in $\chi_2$ and $n_z$ for brevity. Combine the two rotations, after straightforward algebra, we find the effective Hamiltonian $\mathscr{H}_{\pm}$ of the periodically kicked system. It has the form of spin 1/2 in fictitious magnetic field ${\mathbf{B}}_\pm$,
\[
 U_\pm(\mathbf{k},T) = e^{-i \mathscr{H}_{\pm} T} \equiv  e^{-i (\mathbf{B}_{\pm} \cdot \boldsymbol{\sigma})T}.
\]
The magnitude of $\mathbf{B_\pm} $ is nothing but the quasienergy, $\omega_\pm=B_\pm$ (the negative branch of the quasienergy spectrum is simply  $-\omega_\pm$ due to particle-hole symmetry), and it is given by equation
\begin{equation}
\cos ( \omega_\pm T) =  \cos(\chi_1) \cos ( \chi_2)- n_z \sin ( \chi_1) \sin (\chi_2).
\label{qe}
\end{equation}
Explicitly, vector $\mathbf{B_\pm}$ is given by
\begin{equation}
\mathbf{B}_{\pm}T = \frac{ \omega_\pm T }{ \sin (\omega_{\pm} T )} [ \sin \chi_1 ( n_x \cos \chi_2  - n_y \sin \chi_2) \hat{x} + \sin \chi_1 ( n_x \sin \chi_2 + n_y \cos \chi_2) \hat{y}+ ( n_z \sin \chi_1 \cos \chi_2   +\sin \chi_2 \cos \chi_1) \hat{z}].
\end{equation}
Now that each quasienergy band  $\omega_\pm(\mathbf{k})$ is mapped to a spin 1/2 problem, its Chern number can be obtained for example by plotting $\mathbf{B}_{\pm} (\mathbf{k})$ for $\mathbf{k}$ inside the Brillouin zone.

Note that the effective magnetic field $\mathbf{B_\pm}$ is well behaved as $\omega_{\pm}T \rightarrow 0$ but diverges as $\omega_{\pm}T \rightarrow \pi$. Furthermore, we note that 
in the fast driving limit of $T \rightarrow 0$, we recover the static model discussed in the previous section. To see this, we expand the sines and cosines in Eq. (\ref{qe}) to the order of $T^2$ to obtain
\[
\omega_{\pm} T  \simeq  \sqrt{\chi_1^2+\chi_2^2+ \chi_1 \chi_2 n_z}
\]
which agrees with the quasienergy spectrum of the static model given by Eq. (\ref{Estatic}). 

\begin{figure}[h]
\centering
\includegraphics[width=4 in]{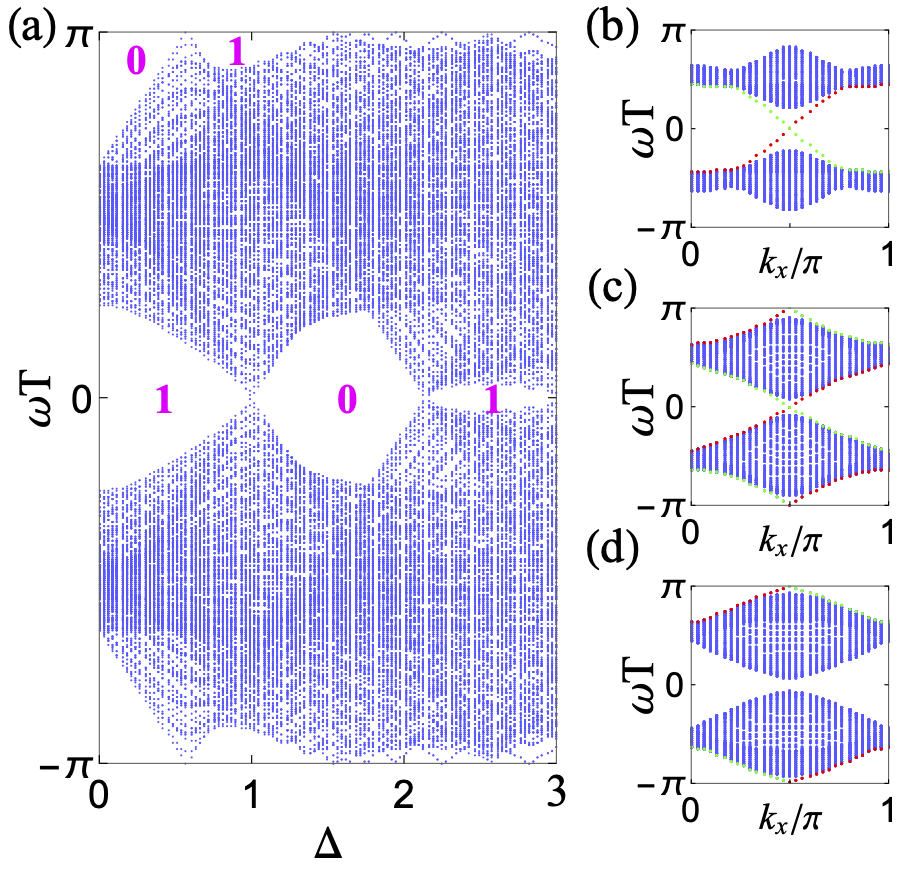}
\caption{Floquet phases of the periodically kicked model. Panel (a) show the quasienergy spectrum $\omega_n$ as $\Delta$ is varied with fixed $\lambda=1$, $\alpha=0.2$, and $T=1$. The gaps are marked with the corresponding winding numbers $W^+_0$ and $W^+_{\pi}$ defined in the main text. Panels (b)-(d) show the quasienergy for strip geometry (open $y$-boundary) at $\Delta=0.3$, $0.8$, and $1.1$. The red/green colors label the edge states near $y=1$ and $y=L_y$ edges; $L_y=60$. The blue dots label the bulk bands.}
\label{PD1}
\end{figure}

The various phases of the periodically driven system can be revealed by graphing  the bulk quasienergy spectrum as function of the parameters of the model. For example, Fig. \ref{PD1} shows the quasienergy as the pairing amplitude $\Delta$ is varied for fixed driving period $T$. Compared to the static cases with only two distinct phases, periodic kicking induces a series of Floquet phases, each of which has a gap at quasienergy 0, or $\pi/T$, or both. The differences between these phases are reflected in their edge states. The extra gap at quasienergy $\pi/T$ (knowns as the $\pi$-gap) gives rise to the possibility of new types of chiral Majorana modes that are intrinsically dynamical in nature \cite{PhysRevB.96.155118,PhysRevB.96.195303}. We will refer to the edge modes
inside the $\pi$-gap (zero gap) as $\pi$-modes (zero modes).  For small $\Delta$, there are a pair of chiral Majorana modes traveling along each edge within the zero gap only, similar to the weak pairing phase of the static model, see Fig. \ref{PD1}(a).  By increasing $\Delta$, the quasienergy bands touch each other at the $\pi$-gap at $\Delta\approx 0.6$. This marks a topological transition into a phase which has a pair of chiral Majorana modes at 
the zero gap as well as the $\pi$-gap, as depicted in Fig. \ref{PD1}(b). As noted earlier, $\mathbf{B}_{\pm}$ becomes ill-defined when the $\pi$-gap closes, $\omega_\pm T=\pi$, accompanied by dramatical changes in the $\pi$-edge modes.
With further increase of $\Delta$, another gap closing occurs at $\Delta\approx 1$ where the zero modes disappear. The system then enters into a Floquet phase characterized by $\pi$-modes only, as shown in Fig. \ref{PD1}(c). This Floquet phase with $\pi$-modes differs from the static strong pairing phase at the same $\Delta$ value. Yet another phase takes over for larger $\Delta$, e.g. $\Delta\sim 2.5$, which features chiral Majorana edge modes at the zero gap.

The intermediate phase for $\Delta$ between 0.6 and 1 is a dynamical anomalous phase \cite{rudner_anomalous_2012}. Even though the Chern numbers of the quasienergy bands are zero, both zero and $\pi$-modes are present. This example shows that Chern numbers alone are insufficient to capture the dynamical origin of the edge modes in many Floquet phases. To correctly describe the bulk-edge correspondence in driven systems, we need to introduce dynamical topological invariants \cite{PhysRevB.96.155118,PhysRevB.96.195303}. In our 2D model, the dynamical invariant is the three-winding number directly constructed from the time-evolution operator. The key concept here is the so-called loop unitary operator, also known as the return map \cite{PhysRevB.96.155118}. The unitary evolution $U(t)$ of a periodically driven system can be decomposed into a unitary loop $\tilde{U}(t)$ satisfying $\tilde{U}(0)=\tilde{U}(T)=I$ and the time evolution of a constant effective Hamiltonian $\mathscr{H}$ \cite{PhysRevB.101.155131}. Explicitly, we can define $\mathscr{H}_{\epsilon}=i\log_{\epsilon} U(T)/T$ as well as the return map
\begin{eqnarray}
\tilde{U}_{\epsilon}(t) =U(t)e^{i \mathscr{H}_{\epsilon}t}.\label{return}
\end{eqnarray}
Here the subscript $\epsilon$ labels the quasienergy gap, and $\log_{\epsilon}$ denotes the logarithm with branch cut lying within gap $\epsilon$. It is apparent from Eq. (\ref{return}) that the topology of $U(t)$ is carried by both $\mathscr{H}_{\epsilon}$ and $\tilde{U}_{\epsilon}(t)$. As discussed above, the spectra of $\mathscr{H}_{\epsilon}$ give the quasienergy band structure, and each band carries a Chern number.
The loop unitary $U_{\epsilon}$ defines a mapping from the space of $(\mathbf{k},t)$, which is a three torus $\textrm{T}^3$, to a space of unitary matrices $U(N)$. The corresponding topological invariant is the integer-valued winding number \cite{rudner_anomalous_2012,PhysRevB.96.195303}
\begin{eqnarray}
W_{\epsilon}=\frac{1}{24\pi^2}\int_{\textrm{T}^3}dk_x dk_y dt~\varepsilon^{\mu\nu\rho} 
\textrm{Tr}[(\tilde{U}_{\epsilon}^{-1}\partial_{\mu}\tilde{U}_{\epsilon})(\tilde{U}_{\epsilon}^{-1}\partial_{\nu}\tilde{U}_{\epsilon})(\tilde{U}_{\epsilon}^{-1}\partial_{\rho}\tilde{U}_{\epsilon})],\label{winding}
\end{eqnarray}
where $t\in[0,T]$, repeated indices $\mu,\nu,\rho\in \{k_x,k_y,t\}$ are summed over, and $\varepsilon^{\mu\nu\rho}$ is the Levi-Civita symbol. Geometrically, the three-winding number $W_{\epsilon}$ measures
the total charge of the dynamical singularities of $\tilde{U}_{\epsilon}$ \cite{nathan2015topological}. The stable singularities of $\tilde{U}_{\epsilon}$ take the form of Weyl points, i.e. isolated points in
the space of $(k_x,k_y,t)$ where the eigenphases of $\tilde{U}$, referred to as phase bands, touch each other, similar to those in a Weyl semimetal 
\cite{nathan2015topological, AnatomyofaPeriodicallyDrivenpWaveSuperconductor}.
For an anomalous Floquet phase with finite $W_{\epsilon}$, the dynamical singularities can not be smoothly deformed away to become a trivial phase 
with $W_{\epsilon}=0$.
One can further prove the following relation between the dynamical invariant $W_{\epsilon}$ and the Chern numbers of the Floquet bands
between gap $\epsilon_2$ and $\epsilon_1$ \cite{rudner_anomalous_2012},
\begin{eqnarray}
\sum_{\epsilon_1<\epsilon_n<\epsilon_2} C_{\epsilon_n}=W_{\epsilon_2}-W_{\epsilon_1}.
\label{w-chern}
\end{eqnarray}
The number of chiral edge states within a given gap $\epsilon$ is given by the invariant $W_{\epsilon}$ \cite{rudner_anomalous_2012}.
For phases that feature two gaps at quasienergy zero and $\pi/T$ respectively, the topological invariants $(W_0,W_{\pi})$ completely characterize
the Floquet phase \cite{PhysRevB.96.155118}.
For our problem, due to the $\mathcal{O}$ symmetry, $U$, $\tilde{U}$ and $\mathscr{H}$ can all be decomposed into the direct sum of $2\times 2$ matrices in the eigenspace of $\mathcal{O}$. For example, the calculation of $W^\pm_{\epsilon}$ only involves $U(2)$ matrices. Then we can define $W_{\epsilon}=W^+_{\epsilon}+W^-_{\epsilon}$ for $\epsilon=0,\pi$.

Armed with the knowledge of $W_\epsilon$, let us revisit the phase diagram in Fig. \ref{PD1} from the viewpoint of dynamical invariants. For $0<\Delta<0.6$, $W^\pm_0=1$ and $W^\pm_{\pi}=0$, which is in agreement of having a pair of chiral Majorana edge mode in the zero gap. For $0.6<\Delta<1$, $W^\pm_0=1$ and $W^\pm_{\pi}=1$. Correspondingly, edge modes are visible in both gaps, and the Chern numbers vanish in accordance to Eq. \eqref{w-chern}. And finally for $1<\Delta<1.5$, $W^\pm_0=0$, $W^\pm_{\pi}=1$, so only $\pi$-modes are present. The total Chern number (sum of contributions from both mirror subspaces) for the quasienergy bands below the zero gap are $2$, $0$, and $-2$, respectively, for the three phases above. We can clearly see that even when the Chern number of the bands are zero, edge states can still emerge. Such anomalous behavior is of dynamical origin and has no static counterparts. 

\begin{figure}[h]
\centering
\includegraphics[width=4 in]{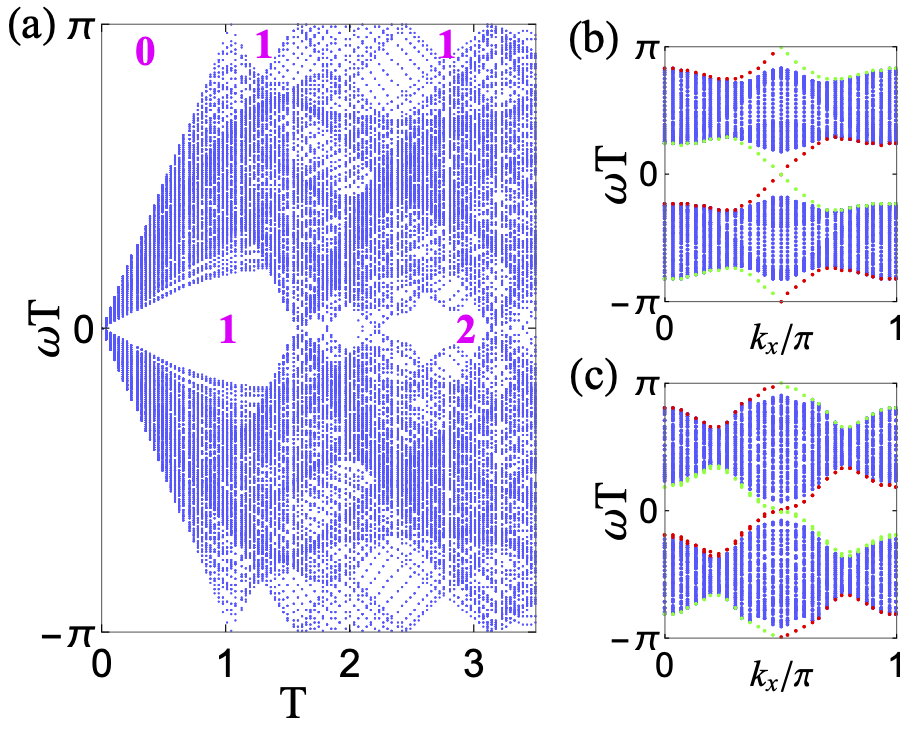}
\caption{The quasienergy spectrum (a) of the periodically kicked model as function of the driving period $T$ with fixed $\lambda=1$, $\alpha=0.2$,  and$\Delta=0.5$. The major gaps are labeled with the corresponding winding numbers $W^+_0$ or $W^+_{\pi}$. Panel b (c) shows the edge state spectrum for strip geometry at $T=1.3$ ($T= 2.8$), confirming the bulk-edge correspondence.}
\label{PD2}
\end{figure}

Fig. \ref{PD2} further illustrates the rich phases of the system for various kicking period $T$. For small $T$, i.e. at high driving frequency, the system behaves like a static system in the weak pairing phase with a pair of chiral Majorana modes within the zero gap. With increasing $T$, the $\pi$-quasienergy gap closes around $T=1$ and then reopens. This ushers in a phase with a pair of chiral Majorana edge modes in the $\pi$-gap, in addition to the edge modes within the zero gap, as depicted in Fig. \ref{PD2}(b). Its dynamical invariants are $W^\pm_0=W^\pm_\pi=1$, the same as the intermediate anomalous phase discussed in the preceding paragraph. Further increasing $T$ leads the system to a gapless phase with no well defined quasienergy gaps. Even further increase in $T$, however, gives rise to a gapped phase around $T=2.8$ with $W^\pm_0=2$ and $W^\pm_\pi=1$. There are in total three pairs of chiral Majorana modes along each edge, two of which are degenerate within the zero gap, as depicted in Fig. \ref{PD2}(c).

\begin{figure}[h]
\centering
\includegraphics[width=4 in]{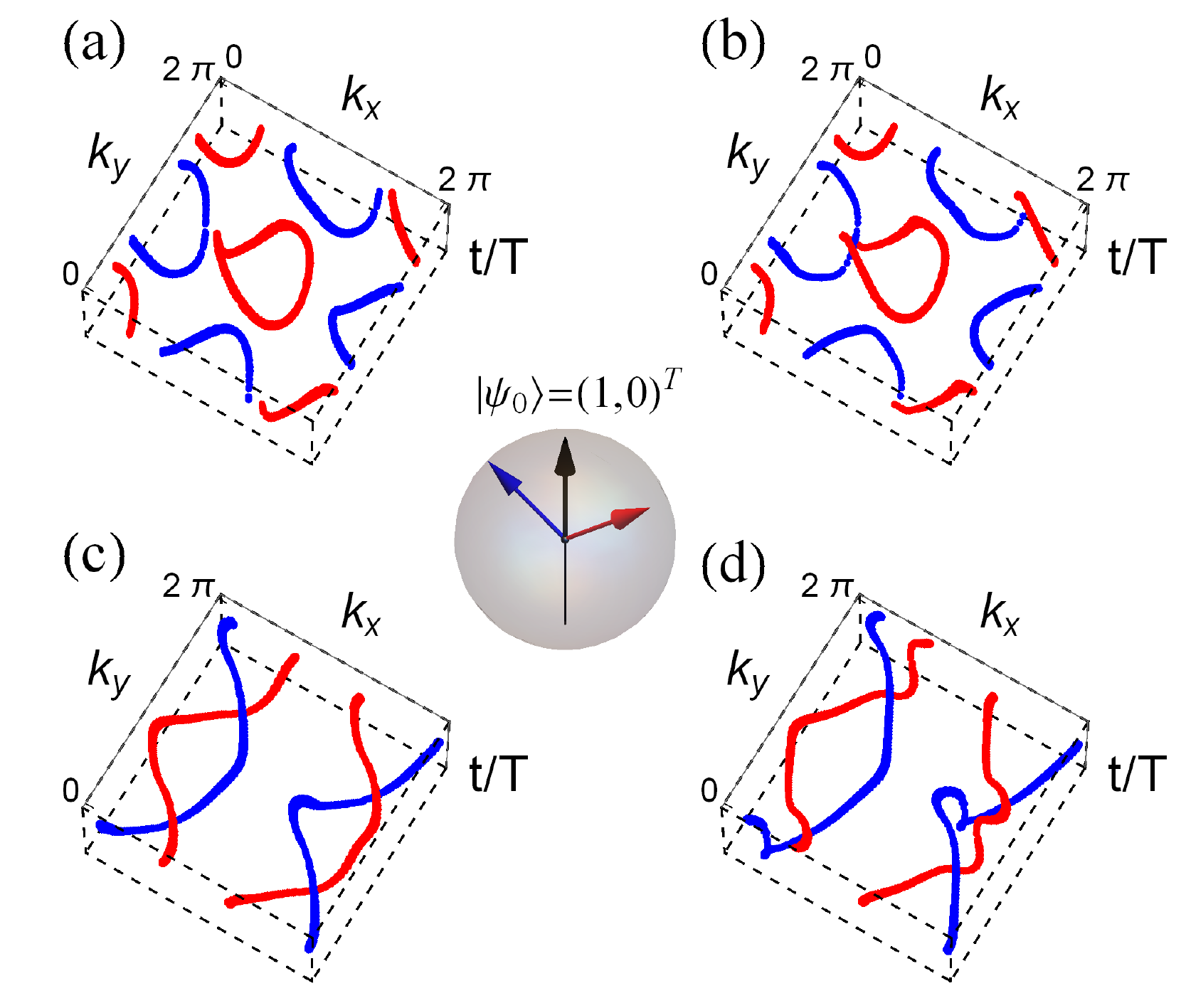}
\caption{Hopf links in the 3D momentum-time space as a characterization of bulk dynamical topology. (a) and (b) show the 
preimages of two chosen target states (red and blue arrows on the Bloch sphere, see the inset) under the evolution of $U^+_0$ and $U^+_\pi$, 
respectively, for $\Delta=0.3$. The Hopf linking number $\mathcal{L}^\pm_0=1$, $\mathcal{L}^\pm_\pi=0$.  (c) and (d) show the results for $\Delta=0.8$. Here the linking number $\mathcal{L}^\pm_0=1$, $\mathcal{L}^\pm_\pi=1$. Note that $k_x$ is plotted from 0 to $2\pi$. The middle inset gives the initial state $|\psi_0\rangle=(1,0)^T$ (black arrow) and two target states $|\psi_{blue}\rangle=\frac{1}{\sqrt{5}}(2,1)^T$ and $|\psi_{red}\rangle=\frac{1}{\sqrt{5}}(-2,1)^T$ (blue and red arrows) on the Bloch sphere. Other parameters are $\lambda=1$, $\alpha=0.2$ and $T=1$.}
\label{fighopf}
\end{figure}

The dynamical topology of the periodically kicked $p$-wave superconductor encoded in the time evolution operator $U_{\epsilon}(t)$ can be further understood from another geometric point of view \cite{PhysRevResearch.1.022003,PhysRevLett.124.160402}. Consider some arbitrary, topologically trivial initial state, for example $|\psi_0(\bm k,t=0)\rangle=(1,0)^T$. The time-evolved state generated by $\tilde{U}_{\epsilon}$ is a spinor,
\begin{eqnarray}
|\psi^\pm_{\epsilon}(\bm k,t)\rangle=\tilde{U}^\pm_{\epsilon}(\bm k,t)|\psi_0(\bm k,t=0)\rangle.
\end{eqnarray} 
It defines a mapping from the three torus T$^3$ to the state space of spinors, the Bloch sphere $S^2$.
This mapping is nothing but the famous Hopf mapping in mathematics, which is characterized by a homotopy invariant $\pi_3(S^2)=\mathbb{Z}$ (we ignore the nontrivial cycles of the torus, which does not play a role here, and treat T$^3$ as three sphere $S^3$) \cite{PhysRevLett.51.2250,PhysRevLett.101.186805,PhysRevB.88.201105}. Explicitly, the Hopf invariant  is defined as follows  \cite{PhysRevResearch.1.022003,PhysRevLett.124.160402,PhysRevLett.118.185701,PhysRevB.101.155131}
\begin{eqnarray}
\mathcal{L}^\pm_{\epsilon}=-\frac{1}{4\pi^2}\int_{\mathrm{T}^3}dk_x dk_y dt~\varepsilon^{\mu\nu\rho}\psi_{\epsilon}^{\dag}\partial_{\mu}\psi_{\epsilon}\partial_{\nu}\psi_{\epsilon}^{\dag}\partial_{\rho}\psi_{\epsilon},
\label{linking}
\end{eqnarray}
with the partial derivative with respective to $k_x$, $k_y$ and $t$ and $\varepsilon^{\mu\nu\rho}$ the Levi-Civita symbol. In Eq. \eqref{linking}, 
we have suppressed the superscript $\pm$ of $\psi$ to avoid clutter. Note that the definition of Hopf invariant is 
possible because the evolution is time periodic, $\tilde{U}_{\epsilon}(t=0)=\tilde{U}_{\epsilon}(t=T)=I$, and 
the $4\times 4$ BdG Hamiltonian can be decomposed into two $2\times 2$ Hamiltonians, $H_\pm$. 

Under the action of the loop unitary $\tilde{U}_{\epsilon}$, for a chosen momentum-time value $(k_x,k_y,t)$, the initial state is transformed to a specific point on the Bloch sphere. Conversely, if we fix one specific point on the Bloch sphere, the preimage of the Hopf mapping will take the form of a closed trajectory  in the 3D momentum-time space (if there is any). If we chose two distinct points on the Bloch sphere (the target states), their preimage trajectories, as two closed loops, will produce a linking pattern in the momentum-time space. The Hopf invariant is nothing but the linking number of the two curves, which is also intimately related to the geometric phase \cite{white1969self,balakrishnan2005gauge}. We note that this equation along with Eq. (\ref{w-chern}) is reminiscent of the CWF theorem in classical differential geometry for twisted space curves \cite{white1969self,balakrishnan2005gauge} whose topology is modeled by a thin ribbon.
One can further show that the linking number of the preimages of any two points on the Bloch sphere is exactly the three-winding number defined earlier in Eq. \eqref{winding} \cite{PhysRevLett.124.160402},
\begin{eqnarray}
\mathcal{L}^\pm_{\epsilon}=W^\pm_{\epsilon}.
\end{eqnarray}
This equivalence provides an intuitive geometric picture of the dynamical topology of our kicked system. It also suggests a promising way for experimental detection of the dynamical topology by measuring the Hopf links, which were recently observed in quench dynamics using momentum-time resolved Bloch-state tomography \cite{tarnowski2019measuring,PhysRevLett.121.250403,yi2019observation}.

The Hopf link characterization of our periodically kicked model is summarized in Fig. \ref{fighopf}. Let us focus on the mirror subspace $+$ (same result applies to the $-$ subspace). An initial state at the north pole on the Bloch sphere will be ``scattered'' to points on the Bloch sphere under the evolution with the loop unitary. At $\Delta=0.3$, the kicked system host edge modes only within the zero gap. Correspondingly, the preimage trajectories of the two distinct target states evolved by $U^+_{\pi}$ are completely disjointed in the momentum-time space as depicted in Fig. \ref{fighopf}(a), and their linking number is $\mathcal{L}^+_{\pi}=0$. In contrast, the preimages of $U^+_{0}$ form links with linking number $\mathcal{L}^+_0=1$ as depicted in Fig. \ref{fighopf}(b). 
At $\Delta=0.8$, the kicked system host both $0$-modes and $\pi$-modes. The system is in the dynamical anomalous phase. The preimage trajectories of both $U^+_{\pi}$ and $U^+_{0}$ are linked together, with linking number $\mathcal{L}^+_0=\mathcal{L}^+_{\pi}=1$, as shown in Fig. \ref{fighopf}(c)(d), in consistent with their bulk winding number $W^+_0=W^+_{\pi}=1$. 
We stress that the time-evolved spinor wavefunction is gauge dependent (recall $H_{BdG}(\mathbf{k})$ depends on the gauge choice), but the linking number is gauge invariant. For our gauge choice, it is most convenient to plot the preimages of time-evolved states in the extended Brillouin zone scheme, for example within the interval $k_x\in [0,2\pi]$ as done in Fig. \ref{fighopf}. In this way, the linking number can be easily counted.

\section{Discussions}

It is well known that periodic driving offers a powerful technique to engineer the band topology of topological insulators. For example, in our previous work, we found a series of Floquet phases for the periodically kicked Harper-Hofstadter at $\pi$ flux (but with no pairing), where the edge modes curiously take the simple form of cosine function in its dispersion \cite{PhysRevB.94.245128}. We have also investigated periodically driven $p$-wave superconductors that consist of
coupled Kitaev chains (but in the absence of magnetic flux or diagonal hopping), where chiral Majorana edge modes were found in the zero gap for a specially designed four-step driving \cite{AnatomyofaPeriodicallyDrivenpWaveSuperconductor}. Compared to these earlier studies, the model introduced here is more general and complex. But paradoxically,  analytical solutions are possible by a consistent decomposition based on symmetry. 
The decomposition also greatly aids the visualization and computation of the Chern numbers and the three-winding numbers. Furthermore, it yields fresh insights by relating the dynamical topology to Hopf links.  The topological invariants computed for each of the various Floquet phases accurately predict the edge modes in the quasienergy gap. 
The main lesson learned from this theoretical toy model is that chiral Majorana edge modes, either static or dynamical (for example the $\pi$-modes), is quite general and not confined to the $p_x\pm ip_y$ state. In particular, the $\pi$-Majorana fermions on the edge are protected by dynamical invariants. This broadens the experimental options for manipulating chiral Majorana fermions for quantum gates or topological quantum computing \cite{lian2018topological,Hu_2019}.
We hope the technical approach presented here is useful for theoretical analysis of multiple-band superconductors that possess additional symmetries, and for visualization of the topological characteristics of other Floquet systems.

This work is supported by AFOSR Grant No. FA9550- 16-1-0006 and NSF Grant No. PHY-1707484.

\bibliographystyle{apsrev4-1}
\bibliography{coldatoms-2020,gases2020}

\begin{thebibliography}{45}%
\makeatletter
\providecommand \@ifxundefined [1]{%
 \@ifx{#1\undefined}
}%
\providecommand \@ifnum [1]{%
 \ifnum #1\expandafter \@firstoftwo
 \else \expandafter \@secondoftwo
 \fi
}%
\providecommand \@ifx [1]{%
 \ifx #1\expandafter \@firstoftwo
 \else \expandafter \@secondoftwo
 \fi
}%
\providecommand \natexlab [1]{#1}%
\providecommand \enquote  [1]{``#1''}%
\providecommand \bibnamefont  [1]{#1}%
\providecommand \bibfnamefont [1]{#1}%
\providecommand \citenamefont [1]{#1}%
\providecommand \href@noop [0]{\@secondoftwo}%
\providecommand \href [0]{\begingroup \@sanitize@url \@href}%
\providecommand \@href[1]{\@@startlink{#1}\@@href}%
\providecommand \@@href[1]{\endgroup#1\@@endlink}%
\providecommand \@sanitize@url [0]{\catcode `\\12\catcode `\$12\catcode
  `\&12\catcode `\#12\catcode `\^12\catcode `\_12\catcode `\%12\relax}%
\providecommand \@@startlink[1]{}%
\providecommand \@@endlink[0]{}%
\providecommand \url  [0]{\begingroup\@sanitize@url \@url }%
\providecommand \@url [1]{\endgroup\@href {#1}{\urlprefix }}%
\providecommand \urlprefix  [0]{URL }%
\providecommand \Eprint [0]{\href }%
\providecommand \doibase [0]{http://dx.doi.org/}%
\providecommand \selectlanguage [0]{\@gobble}%
\providecommand \bibinfo  [0]{\@secondoftwo}%
\providecommand \bibfield  [0]{\@secondoftwo}%
\providecommand \translation [1]{[#1]}%
\providecommand \BibitemOpen [0]{}%
\providecommand \bibitemStop [0]{}%
\providecommand \bibitemNoStop [0]{.\EOS\space}%
\providecommand \EOS [0]{\spacefactor3000\relax}%
\providecommand \BibitemShut  [1]{\csname bibitem#1\endcsname}%
\let\auto@bib@innerbib\@empty
\bibitem [{\citenamefont {Hasan}\ and\ \citenamefont {Kane}(2010)}]{rmp}%
  \BibitemOpen
  \bibfield  {author} {\bibinfo {author} {\bibfnamefont {M.~Z.}\ \bibnamefont
  {Hasan}}\ and\ \bibinfo {author} {\bibfnamefont {C.~L.}\ \bibnamefont
  {Kane}},\ }\href {https://doi.org/10.1103/RevModPhys.82.3045} {\bibfield
  {journal} {\bibinfo  {journal} {Rev. Mod. Phys.}\ }\textbf {\bibinfo {volume}
  {82}},\ \bibinfo {pages} {3045} (\bibinfo {year} {2010})}\BibitemShut
  {NoStop}%
\bibitem [{\citenamefont {Qi}\ and\ \citenamefont {Zhang}(2011)}]{Qi2011}%
  \BibitemOpen
  \bibfield  {author} {\bibinfo {author} {\bibfnamefont {X.-L.}\ \bibnamefont
  {Qi}}\ and\ \bibinfo {author} {\bibfnamefont {S.-C.}\ \bibnamefont {Zhang}},\
  }\href {\doibase 10.1103/RevModPhys.83.1057} {\bibfield  {journal} {\bibinfo
  {journal} {Rev. Mod. Phys.}\ }\textbf {\bibinfo {volume} {83}},\ \bibinfo
  {pages} {1057} (\bibinfo {year} {2011})}\BibitemShut {NoStop}%
\bibitem [{\citenamefont {Kitaev}(2009)}]{kitaev}%
  \BibitemOpen
  \bibfield  {author} {\bibinfo {author} {\bibfnamefont {A.}~\bibnamefont
  {Kitaev}},\ }\href {https://doi.org/10.1063/1.3149495} {\bibfield  {journal}
  {\bibinfo  {journal} {AIP Conf. Proc.}\ }\textbf {\bibinfo {volume} {1134}},\
  \bibinfo {pages} {22} (\bibinfo {year} {2009})}\BibitemShut {NoStop}%
\bibitem [{\citenamefont {Schnyder}\ \emph {et~al.}(2008)\citenamefont
  {Schnyder}, \citenamefont {Ryu}, \citenamefont {Furusaki},\ and\
  \citenamefont {Ludwig}}]{ten-fold}%
  \BibitemOpen
  \bibfield  {author} {\bibinfo {author} {\bibfnamefont {A.~P.}\ \bibnamefont
  {Schnyder}}, \bibinfo {author} {\bibfnamefont {S.}~\bibnamefont {Ryu}},
  \bibinfo {author} {\bibfnamefont {A.}~\bibnamefont {Furusaki}}, \ and\
  \bibinfo {author} {\bibfnamefont {A.~W.~W.}\ \bibnamefont {Ludwig}},\ }\href
  {\doibase 10.1103/PhysRevB.78.195125} {\bibfield  {journal} {\bibinfo
  {journal} {Phys. Rev. B}\ }\textbf {\bibinfo {volume} {78}},\ \bibinfo
  {pages} {195125} (\bibinfo {year} {2008})}\BibitemShut {NoStop}%
\bibitem [{\citenamefont {Read}\ and\ \citenamefont
  {Green}(2000)}]{PhysRevB.61.10267}%
  \BibitemOpen
  \bibfield  {author} {\bibinfo {author} {\bibfnamefont {N.}~\bibnamefont
  {Read}}\ and\ \bibinfo {author} {\bibfnamefont {D.}~\bibnamefont {Green}},\
  }\href {\doibase 10.1103/PhysRevB.61.10267} {\bibfield  {journal} {\bibinfo
  {journal} {Phys. Rev. B}\ }\textbf {\bibinfo {volume} {61}},\ \bibinfo
  {pages} {10267} (\bibinfo {year} {2000})}\BibitemShut {NoStop}%
\bibitem [{\citenamefont {Ivanov}(2001)}]{PhysRevLett.86.268}%
  \BibitemOpen
  \bibfield  {author} {\bibinfo {author} {\bibfnamefont {D.~A.}\ \bibnamefont
  {Ivanov}},\ }\href {\doibase 10.1103/PhysRevLett.86.268} {\bibfield
  {journal} {\bibinfo  {journal} {Phys. Rev. Lett.}\ }\textbf {\bibinfo
  {volume} {86}},\ \bibinfo {pages} {268} (\bibinfo {year} {2001})}\BibitemShut
  {NoStop}%
\bibitem [{\citenamefont {Nayak}\ \emph {et~al.}(2008)\citenamefont {Nayak},
  \citenamefont {Simon}, \citenamefont {Stern}, \citenamefont {Freedman},\ and\
  \citenamefont {Das~Sarma}}]{RevModPhys.80.1083}%
  \BibitemOpen
  \bibfield  {author} {\bibinfo {author} {\bibfnamefont {C.}~\bibnamefont
  {Nayak}}, \bibinfo {author} {\bibfnamefont {S.~H.}\ \bibnamefont {Simon}},
  \bibinfo {author} {\bibfnamefont {A.}~\bibnamefont {Stern}}, \bibinfo
  {author} {\bibfnamefont {M.}~\bibnamefont {Freedman}}, \ and\ \bibinfo
  {author} {\bibfnamefont {S.}~\bibnamefont {Das~Sarma}},\ }\href {\doibase
  10.1103/RevModPhys.80.1083} {\bibfield  {journal} {\bibinfo  {journal} {Rev.
  Mod. Phys.}\ }\textbf {\bibinfo {volume} {80}},\ \bibinfo {pages} {1083}
  (\bibinfo {year} {2008})}\BibitemShut {NoStop}%
\bibitem [{\citenamefont {Harper}(1955)}]{harper1955single}%
  \BibitemOpen
  \bibfield  {author} {\bibinfo {author} {\bibfnamefont {P.~G.}\ \bibnamefont
  {Harper}},\ }\href@noop {} {\bibfield  {journal} {\bibinfo  {journal}
  {Proceedings of the Physical Society. Section A}\ }\textbf {\bibinfo {volume}
  {68}},\ \bibinfo {pages} {874} (\bibinfo {year} {1955})}\BibitemShut
  {NoStop}%
\bibitem [{\citenamefont {Hofstadter}(1976)}]{hofstadter_energy_1976}%
  \BibitemOpen
  \bibfield  {author} {\bibinfo {author} {\bibfnamefont {D.~R.}\ \bibnamefont
  {Hofstadter}},\ }\href {\doibase 10.1103/PhysRevB.14.2239} {\bibfield
  {journal} {\bibinfo  {journal} {Phys. Rev. B}\ }\textbf {\bibinfo {volume}
  {14}},\ \bibinfo {pages} {2239} (\bibinfo {year} {1976})}\BibitemShut
  {NoStop}%
\bibitem [{\citenamefont {Kitaev}(2001)}]{kitaev2001unpaired}%
  \BibitemOpen
  \bibfield  {author} {\bibinfo {author} {\bibfnamefont {A.~Y.}\ \bibnamefont
  {Kitaev}},\ }\href
  {https://iopscience.iop.org/article/10.1070/1063-7869/44/10S/S29/pdf}
  {\bibfield  {journal} {\bibinfo  {journal} {Physics-Uspekhi}\ }\textbf
  {\bibinfo {volume} {44}},\ \bibinfo {pages} {131} (\bibinfo {year}
  {2001})}\BibitemShut {NoStop}%
\bibitem [{\citenamefont {Dalibard}\ \emph {et~al.}(2011)\citenamefont
  {Dalibard}, \citenamefont {Gerbier}, \citenamefont
  {Juzeli\ifmmode~\bar{u}\else \={u}\fi{}nas},\ and\ \citenamefont
  {\"Ohberg}}]{gauge-rev}%
  \BibitemOpen
  \bibfield  {author} {\bibinfo {author} {\bibfnamefont {J.}~\bibnamefont
  {Dalibard}}, \bibinfo {author} {\bibfnamefont {F.}~\bibnamefont {Gerbier}},
  \bibinfo {author} {\bibfnamefont {G.}~\bibnamefont
  {Juzeli\ifmmode~\bar{u}\else \={u}\fi{}nas}}, \ and\ \bibinfo {author}
  {\bibfnamefont {P.}~\bibnamefont {\"Ohberg}},\ }\href {\doibase
  10.1103/RevModPhys.83.1523} {\bibfield  {journal} {\bibinfo  {journal} {Rev.
  Mod. Phys.}\ }\textbf {\bibinfo {volume} {83}},\ \bibinfo {pages} {1523}
  (\bibinfo {year} {2011})}\BibitemShut {NoStop}%
\bibitem [{\citenamefont {Thouless}(1983)}]{PhysRevB.28.4272}%
  \BibitemOpen
  \bibfield  {author} {\bibinfo {author} {\bibfnamefont {D.~J.}\ \bibnamefont
  {Thouless}},\ }\href {\doibase 10.1103/PhysRevB.28.4272} {\bibfield
  {journal} {\bibinfo  {journal} {Phys. Rev. B}\ }\textbf {\bibinfo {volume}
  {28}},\ \bibinfo {pages} {4272} (\bibinfo {year} {1983})}\BibitemShut
  {NoStop}%
\bibitem [{\citenamefont {Hatsugai}\ and\ \citenamefont
  {Kohmoto}(1990)}]{PhysRevB.42.8282}%
  \BibitemOpen
  \bibfield  {author} {\bibinfo {author} {\bibfnamefont {Y.}~\bibnamefont
  {Hatsugai}}\ and\ \bibinfo {author} {\bibfnamefont {M.}~\bibnamefont
  {Kohmoto}},\ }\href {\doibase 10.1103/PhysRevB.42.8282} {\bibfield  {journal}
  {\bibinfo  {journal} {Phys. Rev. B}\ }\textbf {\bibinfo {volume} {42}},\
  \bibinfo {pages} {8282} (\bibinfo {year} {1990})}\BibitemShut {NoStop}%
\bibitem [{\citenamefont {Lababidi}\ \emph {et~al.}(2014)\citenamefont
  {Lababidi}, \citenamefont {Satija},\ and\ \citenamefont
  {Zhao}}]{PhysRevLett.112.026805}%
  \BibitemOpen
  \bibfield  {author} {\bibinfo {author} {\bibfnamefont {M.}~\bibnamefont
  {Lababidi}}, \bibinfo {author} {\bibfnamefont {I.~I.}\ \bibnamefont
  {Satija}}, \ and\ \bibinfo {author} {\bibfnamefont {E.}~\bibnamefont
  {Zhao}},\ }\href {\doibase 10.1103/PhysRevLett.112.026805} {\bibfield
  {journal} {\bibinfo  {journal} {Phys. Rev. Lett.}\ }\textbf {\bibinfo
  {volume} {112}},\ \bibinfo {pages} {026805} (\bibinfo {year}
  {2014})}\BibitemShut {NoStop}%
\bibitem [{\citenamefont {Bukov}\ \emph {et~al.}(2015)\citenamefont {Bukov},
  \citenamefont {D'Alessio},\ and\ \citenamefont
  {Polkovnikov}}]{bukov2015universal}%
  \BibitemOpen
  \bibfield  {author} {\bibinfo {author} {\bibfnamefont {M.}~\bibnamefont
  {Bukov}}, \bibinfo {author} {\bibfnamefont {L.}~\bibnamefont {D'Alessio}}, \
  and\ \bibinfo {author} {\bibfnamefont {A.}~\bibnamefont {Polkovnikov}},\
  }\href {\doibase 10.1080/00018732.2015.1055918} {\bibfield  {journal}
  {\bibinfo  {journal} {Advances in Physics}\ }\textbf {\bibinfo {volume}
  {64}},\ \bibinfo {pages} {139} (\bibinfo {year} {2015})}\BibitemShut
  {NoStop}%
\bibitem [{\citenamefont {Eckardt}(2017)}]{RevModPhys.89.011004}%
  \BibitemOpen
  \bibfield  {author} {\bibinfo {author} {\bibfnamefont {A.}~\bibnamefont
  {Eckardt}},\ }\href {\doibase 10.1103/RevModPhys.89.011004} {\bibfield
  {journal} {\bibinfo  {journal} {Rev. Mod. Phys.}\ }\textbf {\bibinfo {volume}
  {89}},\ \bibinfo {pages} {011004} (\bibinfo {year} {2017})}\BibitemShut
  {NoStop}%
\bibitem [{\citenamefont {Zhang}\ \emph {et~al.}(2013)\citenamefont {Zhang},
  \citenamefont {Kane},\ and\ \citenamefont {Mele}}]{PhysRevLett.111.056403}%
  \BibitemOpen
  \bibfield  {author} {\bibinfo {author} {\bibfnamefont {F.}~\bibnamefont
  {Zhang}}, \bibinfo {author} {\bibfnamefont {C.~L.}\ \bibnamefont {Kane}}, \
  and\ \bibinfo {author} {\bibfnamefont {E.~J.}\ \bibnamefont {Mele}},\ }\href
  {\doibase 10.1103/PhysRevLett.111.056403} {\bibfield  {journal} {\bibinfo
  {journal} {Phys. Rev. Lett.}\ }\textbf {\bibinfo {volume} {111}},\ \bibinfo
  {pages} {056403} (\bibinfo {year} {2013})}\BibitemShut {NoStop}%
\bibitem [{\citenamefont {Hu}\ \emph {et~al.}(2019)\citenamefont {Hu},
  \citenamefont {Satija},\ and\ \citenamefont {Zhao}}]{Hu_2019}%
  \BibitemOpen
  \bibfield  {author} {\bibinfo {author} {\bibfnamefont {H.}~\bibnamefont
  {Hu}}, \bibinfo {author} {\bibfnamefont {I.~I.}\ \bibnamefont {Satija}}, \
  and\ \bibinfo {author} {\bibfnamefont {E.}~\bibnamefont {Zhao}},\ }\href
  {\doibase 10.1088/1367-2630/ab5cad} {\bibfield  {journal} {\bibinfo
  {journal} {New Journal of Physics}\ }\textbf {\bibinfo {volume} {21}},\
  \bibinfo {pages} {123014} (\bibinfo {year} {2019})}\BibitemShut {NoStop}%
\bibitem [{\citenamefont {Shiozaki}\ and\ \citenamefont
  {Sato}(2014)}]{PhysRevB.90.165114}%
  \BibitemOpen
  \bibfield  {author} {\bibinfo {author} {\bibfnamefont {K.}~\bibnamefont
  {Shiozaki}}\ and\ \bibinfo {author} {\bibfnamefont {M.}~\bibnamefont
  {Sato}},\ }\href {\doibase 10.1103/PhysRevB.90.165114} {\bibfield  {journal}
  {\bibinfo  {journal} {Phys. Rev. B}\ }\textbf {\bibinfo {volume} {90}},\
  \bibinfo {pages} {165114} (\bibinfo {year} {2014})}\BibitemShut {NoStop}%
\bibitem [{\citenamefont {Oka}\ and\ \citenamefont
  {Aoki}(2009)}]{PhysRevB.79.081406}%
  \BibitemOpen
  \bibfield  {author} {\bibinfo {author} {\bibfnamefont {T.}~\bibnamefont
  {Oka}}\ and\ \bibinfo {author} {\bibfnamefont {H.}~\bibnamefont {Aoki}},\
  }\href {\doibase 10.1103/PhysRevB.79.081406} {\bibfield  {journal} {\bibinfo
  {journal} {Phys. Rev. B}\ }\textbf {\bibinfo {volume} {79}},\ \bibinfo
  {pages} {081406} (\bibinfo {year} {2009})}\BibitemShut {NoStop}%
\bibitem [{\citenamefont {Lindner}\ \emph
  {et~al.}(2011{\natexlab{a}})\citenamefont {Lindner}, \citenamefont {Refael},\
  and\ \citenamefont {Galitski}}]{lindner_floquet_2011}%
  \BibitemOpen
  \bibfield  {author} {\bibinfo {author} {\bibfnamefont {N.~H.}\ \bibnamefont
  {Lindner}}, \bibinfo {author} {\bibfnamefont {G.}~\bibnamefont {Refael}}, \
  and\ \bibinfo {author} {\bibfnamefont {V.}~\bibnamefont {Galitski}},\ }\href
  {\doibase 10.1038/nphys1926} {\bibfield  {journal} {\bibinfo  {journal}
  {Nature Physics}\ }\textbf {\bibinfo {volume} {7}},\ \bibinfo {pages} {490}
  (\bibinfo {year} {2011}{\natexlab{a}})}\BibitemShut {NoStop}%
\bibitem [{\citenamefont {Kitagawa}\ \emph {et~al.}(2010)\citenamefont
  {Kitagawa}, \citenamefont {Berg}, \citenamefont {Rudner},\ and\ \citenamefont
  {Demler}}]{kitagawa_topological_2010}%
  \BibitemOpen
  \bibfield  {author} {\bibinfo {author} {\bibfnamefont {T.}~\bibnamefont
  {Kitagawa}}, \bibinfo {author} {\bibfnamefont {E.}~\bibnamefont {Berg}},
  \bibinfo {author} {\bibfnamefont {M.}~\bibnamefont {Rudner}}, \ and\ \bibinfo
  {author} {\bibfnamefont {E.}~\bibnamefont {Demler}},\ }\href {\doibase
  10.1103/PhysRevB.82.235114} {\bibfield  {journal} {\bibinfo  {journal} {Phys.
  Rev. B}\ }\textbf {\bibinfo {volume} {82}},\ \bibinfo {pages} {235114}
  (\bibinfo {year} {2010})}\BibitemShut {NoStop}%
\bibitem [{\citenamefont {Lindner}\ \emph
  {et~al.}(2011{\natexlab{b}})\citenamefont {Lindner}, \citenamefont {Bergman},
  \citenamefont {Refael},\ and\ \citenamefont
  {Galitski}}]{lindner_topological_2011}%
  \BibitemOpen
  \bibfield  {author} {\bibinfo {author} {\bibfnamefont {N.~H.}\ \bibnamefont
  {Lindner}}, \bibinfo {author} {\bibfnamefont {D.~L.}\ \bibnamefont
  {Bergman}}, \bibinfo {author} {\bibfnamefont {G.}~\bibnamefont {Refael}}, \
  and\ \bibinfo {author} {\bibfnamefont {V.}~\bibnamefont {Galitski}},\ }\href
  {http://arxiv.org/abs/1111.4518} {\bibfield  {journal} {\bibinfo  {journal}
  {{arXiv:1111.4518}}\ } (\bibinfo {year} {2011}{\natexlab{b}})}\BibitemShut
  {NoStop}%
\bibitem [{\citenamefont {Kitagawa}\ \emph {et~al.}(2011)\citenamefont
  {Kitagawa}, \citenamefont {Oka}, \citenamefont {Brataas}, \citenamefont
  {Fu},\ and\ \citenamefont {Demler}}]{kitagawa_transport_2011}%
  \BibitemOpen
  \bibfield  {author} {\bibinfo {author} {\bibfnamefont {T.}~\bibnamefont
  {Kitagawa}}, \bibinfo {author} {\bibfnamefont {T.}~\bibnamefont {Oka}},
  \bibinfo {author} {\bibfnamefont {A.}~\bibnamefont {Brataas}}, \bibinfo
  {author} {\bibfnamefont {L.}~\bibnamefont {Fu}}, \ and\ \bibinfo {author}
  {\bibfnamefont {E.}~\bibnamefont {Demler}},\ }\href {\doibase
  10.1103/PhysRevB.84.235108} {\bibfield  {journal} {\bibinfo  {journal} {Phys.
  Rev. B}\ }\textbf {\bibinfo {volume} {84}},\ \bibinfo {pages} {235108}
  (\bibinfo {year} {2011})}\BibitemShut {NoStop}%
\bibitem [{\citenamefont {Jiang}\ \emph {et~al.}(2011)\citenamefont {Jiang},
  \citenamefont {Kitagawa}, \citenamefont {Alicea}, \citenamefont {Akhmerov},
  \citenamefont {Pekker}, \citenamefont {Refael}, \citenamefont {Cirac},
  \citenamefont {Demler}, \citenamefont {Lukin},\ and\ \citenamefont
  {Zoller}}]{jiang_majorana_2011}%
  \BibitemOpen
  \bibfield  {author} {\bibinfo {author} {\bibfnamefont {L.}~\bibnamefont
  {Jiang}}, \bibinfo {author} {\bibfnamefont {T.}~\bibnamefont {Kitagawa}},
  \bibinfo {author} {\bibfnamefont {J.}~\bibnamefont {Alicea}}, \bibinfo
  {author} {\bibfnamefont {A.~R.}\ \bibnamefont {Akhmerov}}, \bibinfo {author}
  {\bibfnamefont {D.}~\bibnamefont {Pekker}}, \bibinfo {author} {\bibfnamefont
  {G.}~\bibnamefont {Refael}}, \bibinfo {author} {\bibfnamefont {J.~I.}\
  \bibnamefont {Cirac}}, \bibinfo {author} {\bibfnamefont {E.}~\bibnamefont
  {Demler}}, \bibinfo {author} {\bibfnamefont {M.~D.}\ \bibnamefont {Lukin}}, \
  and\ \bibinfo {author} {\bibfnamefont {P.}~\bibnamefont {Zoller}},\ }\href
  {\doibase 10.1103/PhysRevLett.106.220402} {\bibfield  {journal} {\bibinfo
  {journal} {Phys. Rev. Lett.}\ }\textbf {\bibinfo {volume} {106}},\ \bibinfo
  {pages} {220402} (\bibinfo {year} {2011})}\BibitemShut {NoStop}%
\bibitem [{\citenamefont {Gu}\ \emph {et~al.}(2011)\citenamefont {Gu},
  \citenamefont {Fertig}, \citenamefont {Arovas},\ and\ \citenamefont
  {Auerbach}}]{gu_floquet_2011}%
  \BibitemOpen
  \bibfield  {author} {\bibinfo {author} {\bibfnamefont {Z.}~\bibnamefont
  {Gu}}, \bibinfo {author} {\bibfnamefont {H.~A.}\ \bibnamefont {Fertig}},
  \bibinfo {author} {\bibfnamefont {D.~P.}\ \bibnamefont {Arovas}}, \ and\
  \bibinfo {author} {\bibfnamefont {A.}~\bibnamefont {Auerbach}},\ }\href
  {\doibase 10.1103/PhysRevLett.107.216601} {\bibfield  {journal} {\bibinfo
  {journal} {Phys. Rev. Lett.}\ }\textbf {\bibinfo {volume} {107}},\ \bibinfo
  {pages} {216601} (\bibinfo {year} {2011})}\BibitemShut {NoStop}%
\bibitem [{\citenamefont {Rudner}\ \emph {et~al.}(2013)\citenamefont {Rudner},
  \citenamefont {Lindner}, \citenamefont {Berg},\ and\ \citenamefont
  {Levin}}]{rudner_anomalous_2012}%
  \BibitemOpen
  \bibfield  {author} {\bibinfo {author} {\bibfnamefont {M.~S.}\ \bibnamefont
  {Rudner}}, \bibinfo {author} {\bibfnamefont {N.~H.}\ \bibnamefont {Lindner}},
  \bibinfo {author} {\bibfnamefont {E.}~\bibnamefont {Berg}}, \ and\ \bibinfo
  {author} {\bibfnamefont {M.}~\bibnamefont {Levin}},\ }\href {\doibase
  10.1103/PhysRevX.3.031005} {\bibfield  {journal} {\bibinfo  {journal} {Phys.
  Rev. X}\ }\textbf {\bibinfo {volume} {3}},\ \bibinfo {pages} {031005}
  (\bibinfo {year} {2013})}\BibitemShut {NoStop}%
\bibitem [{\citenamefont {Roy}\ and\ \citenamefont
  {Harper}(2017)}]{PhysRevB.96.155118}%
  \BibitemOpen
  \bibfield  {author} {\bibinfo {author} {\bibfnamefont {R.}~\bibnamefont
  {Roy}}\ and\ \bibinfo {author} {\bibfnamefont {F.}~\bibnamefont {Harper}},\
  }\href {\doibase 10.1103/PhysRevB.96.155118} {\bibfield  {journal} {\bibinfo
  {journal} {Phys. Rev. B}\ }\textbf {\bibinfo {volume} {96}},\ \bibinfo
  {pages} {155118} (\bibinfo {year} {2017})}\BibitemShut {NoStop}%
\bibitem [{\citenamefont {Yao}\ \emph {et~al.}(2017)\citenamefont {Yao},
  \citenamefont {Yan},\ and\ \citenamefont {Wang}}]{PhysRevB.96.195303}%
  \BibitemOpen
  \bibfield  {author} {\bibinfo {author} {\bibfnamefont {S.}~\bibnamefont
  {Yao}}, \bibinfo {author} {\bibfnamefont {Z.}~\bibnamefont {Yan}}, \ and\
  \bibinfo {author} {\bibfnamefont {Z.}~\bibnamefont {Wang}},\ }\href {\doibase
  10.1103/PhysRevB.96.195303} {\bibfield  {journal} {\bibinfo  {journal} {Phys.
  Rev. B}\ }\textbf {\bibinfo {volume} {96}},\ \bibinfo {pages} {195303}
  (\bibinfo {year} {2017})}\BibitemShut {NoStop}%
\bibitem [{\citenamefont {Hu}\ \emph {et~al.}(2020)\citenamefont {Hu},
  \citenamefont {Yang},\ and\ \citenamefont {Zhao}}]{PhysRevB.101.155131}%
  \BibitemOpen
  \bibfield  {author} {\bibinfo {author} {\bibfnamefont {H.}~\bibnamefont
  {Hu}}, \bibinfo {author} {\bibfnamefont {C.}~\bibnamefont {Yang}}, \ and\
  \bibinfo {author} {\bibfnamefont {E.}~\bibnamefont {Zhao}},\ }\href {\doibase
  10.1103/PhysRevB.101.155131} {\bibfield  {journal} {\bibinfo  {journal}
  {Phys. Rev. B}\ }\textbf {\bibinfo {volume} {101}},\ \bibinfo {pages}
  {155131} (\bibinfo {year} {2020})}\BibitemShut {NoStop}%
\bibitem [{\citenamefont {Nathan}\ and\ \citenamefont
  {Rudner}(2015)}]{nathan2015topological}%
  \BibitemOpen
  \bibfield  {author} {\bibinfo {author} {\bibfnamefont {F.}~\bibnamefont
  {Nathan}}\ and\ \bibinfo {author} {\bibfnamefont {M.~S.}\ \bibnamefont
  {Rudner}},\ }\href {\doibase 10.1088/1367-2630/17/12/125014} {\bibfield
  {journal} {\bibinfo  {journal} {New Journal of Physics}\ }\textbf {\bibinfo
  {volume} {17}},\ \bibinfo {pages} {125014} (\bibinfo {year}
  {2015})}\BibitemShut {NoStop}%
\bibitem [{\citenamefont
  {Zhao}(2016)}]{AnatomyofaPeriodicallyDrivenpWaveSuperconductor}%
  \BibitemOpen
  \bibfield  {author} {\bibinfo {author} {\bibfnamefont {E.}~\bibnamefont
  {Zhao}},\ }\href {\doibase https://doi.org/10.1515/zna-2016-0074} {\bibfield
  {journal} {\bibinfo  {journal} {Zeitschrift f{\"u}r Naturforschung A}\
  }\textbf {\bibinfo {volume} {71}},\ \bibinfo {pages} {883 } (\bibinfo {year}
  {2016})}\BibitemShut {NoStop}%
\bibitem [{\citenamefont {\"Unal}\ \emph {et~al.}(2019)\citenamefont {\"Unal},
  \citenamefont {Eckardt},\ and\ \citenamefont
  {Slager}}]{PhysRevResearch.1.022003}%
  \BibitemOpen
  \bibfield  {author} {\bibinfo {author} {\bibfnamefont {F.~N.}\ \bibnamefont
  {\"Unal}}, \bibinfo {author} {\bibfnamefont {A.}~\bibnamefont {Eckardt}}, \
  and\ \bibinfo {author} {\bibfnamefont {R.-J.}\ \bibnamefont {Slager}},\
  }\href {\doibase 10.1103/PhysRevResearch.1.022003} {\bibfield  {journal}
  {\bibinfo  {journal} {Phys. Rev. Research}\ }\textbf {\bibinfo {volume}
  {1}},\ \bibinfo {pages} {022003} (\bibinfo {year} {2019})}\BibitemShut
  {NoStop}%
\bibitem [{\citenamefont {Hu}\ and\ \citenamefont
  {Zhao}(2020)}]{PhysRevLett.124.160402}%
  \BibitemOpen
  \bibfield  {author} {\bibinfo {author} {\bibfnamefont {H.}~\bibnamefont
  {Hu}}\ and\ \bibinfo {author} {\bibfnamefont {E.}~\bibnamefont {Zhao}},\
  }\href {\doibase 10.1103/PhysRevLett.124.160402} {\bibfield  {journal}
  {\bibinfo  {journal} {Phys. Rev. Lett.}\ }\textbf {\bibinfo {volume} {124}},\
  \bibinfo {pages} {160402} (\bibinfo {year} {2020})}\BibitemShut {NoStop}%
\bibitem [{\citenamefont {Wilczek}\ and\ \citenamefont
  {Zee}(1983)}]{PhysRevLett.51.2250}%
  \BibitemOpen
  \bibfield  {author} {\bibinfo {author} {\bibfnamefont {F.}~\bibnamefont
  {Wilczek}}\ and\ \bibinfo {author} {\bibfnamefont {A.}~\bibnamefont {Zee}},\
  }\href {\doibase 10.1103/PhysRevLett.51.2250} {\bibfield  {journal} {\bibinfo
   {journal} {Phys. Rev. Lett.}\ }\textbf {\bibinfo {volume} {51}},\ \bibinfo
  {pages} {2250} (\bibinfo {year} {1983})}\BibitemShut {NoStop}%
\bibitem [{\citenamefont {Moore}\ \emph {et~al.}(2008)\citenamefont {Moore},
  \citenamefont {Ran},\ and\ \citenamefont {Wen}}]{PhysRevLett.101.186805}%
  \BibitemOpen
  \bibfield  {author} {\bibinfo {author} {\bibfnamefont {J.~E.}\ \bibnamefont
  {Moore}}, \bibinfo {author} {\bibfnamefont {Y.}~\bibnamefont {Ran}}, \ and\
  \bibinfo {author} {\bibfnamefont {X.-G.}\ \bibnamefont {Wen}},\ }\href
  {\doibase 10.1103/PhysRevLett.101.186805} {\bibfield  {journal} {\bibinfo
  {journal} {Phys. Rev. Lett.}\ }\textbf {\bibinfo {volume} {101}},\ \bibinfo
  {pages} {186805} (\bibinfo {year} {2008})}\BibitemShut {NoStop}%
\bibitem [{\citenamefont {Deng}\ \emph {et~al.}(2013)\citenamefont {Deng},
  \citenamefont {Wang}, \citenamefont {Shen},\ and\ \citenamefont
  {Duan}}]{PhysRevB.88.201105}%
  \BibitemOpen
  \bibfield  {author} {\bibinfo {author} {\bibfnamefont {D.-L.}\ \bibnamefont
  {Deng}}, \bibinfo {author} {\bibfnamefont {S.-T.}\ \bibnamefont {Wang}},
  \bibinfo {author} {\bibfnamefont {C.}~\bibnamefont {Shen}}, \ and\ \bibinfo
  {author} {\bibfnamefont {L.-M.}\ \bibnamefont {Duan}},\ }\href {\doibase
  10.1103/PhysRevB.88.201105} {\bibfield  {journal} {\bibinfo  {journal} {Phys.
  Rev. B}\ }\textbf {\bibinfo {volume} {88}},\ \bibinfo {pages} {201105(R)}
  (\bibinfo {year} {2013})}\BibitemShut {NoStop}%
\bibitem [{\citenamefont {Wang}\ \emph {et~al.}(2017)\citenamefont {Wang},
  \citenamefont {Zhang}, \citenamefont {Chen}, \citenamefont {Yu},\ and\
  \citenamefont {Zhai}}]{PhysRevLett.118.185701}%
  \BibitemOpen
  \bibfield  {author} {\bibinfo {author} {\bibfnamefont {C.}~\bibnamefont
  {Wang}}, \bibinfo {author} {\bibfnamefont {P.}~\bibnamefont {Zhang}},
  \bibinfo {author} {\bibfnamefont {X.}~\bibnamefont {Chen}}, \bibinfo {author}
  {\bibfnamefont {J.}~\bibnamefont {Yu}}, \ and\ \bibinfo {author}
  {\bibfnamefont {H.}~\bibnamefont {Zhai}},\ }\href {\doibase
  10.1103/PhysRevLett.118.185701} {\bibfield  {journal} {\bibinfo  {journal}
  {Phys. Rev. Lett.}\ }\textbf {\bibinfo {volume} {118}},\ \bibinfo {pages}
  {185701} (\bibinfo {year} {2017})}\BibitemShut {NoStop}%
\bibitem [{\citenamefont {White}(1969)}]{white1969self}%
  \BibitemOpen
  \bibfield  {author} {\bibinfo {author} {\bibfnamefont {J.~H.}\ \bibnamefont
  {White}},\ }\href@noop {} {\bibfield  {journal} {\bibinfo  {journal}
  {American journal of mathematics}\ }\textbf {\bibinfo {volume} {91}},\
  \bibinfo {pages} {693} (\bibinfo {year} {1969})}\BibitemShut {NoStop}%
\bibitem [{\citenamefont {Balakrishnan}\ and\ \citenamefont
  {Satija}(2005)}]{balakrishnan2005gauge}%
  \BibitemOpen
  \bibfield  {author} {\bibinfo {author} {\bibfnamefont {R.}~\bibnamefont
  {Balakrishnan}}\ and\ \bibinfo {author} {\bibfnamefont {I.~I.}\ \bibnamefont
  {Satija}},\ }\href {https://arxiv.org/abs/math-ph/0507039} {\bibfield
  {journal} {\bibinfo  {journal} {arXiv:math-ph/0507039}\ } (\bibinfo {year}
  {2005})}\BibitemShut {NoStop}%
\bibitem [{\citenamefont {Tarnowski}\ \emph {et~al.}(2019)\citenamefont
  {Tarnowski}, \citenamefont {{\"U}nal}, \citenamefont {Fl{\"a}schner},
  \citenamefont {Rem}, \citenamefont {Eckardt}, \citenamefont {Sengstock},\
  and\ \citenamefont {Weitenberg}}]{tarnowski2019measuring}%
  \BibitemOpen
  \bibfield  {author} {\bibinfo {author} {\bibfnamefont {M.}~\bibnamefont
  {Tarnowski}}, \bibinfo {author} {\bibfnamefont {F.~N.}\ \bibnamefont
  {{\"U}nal}}, \bibinfo {author} {\bibfnamefont {N.}~\bibnamefont
  {Fl{\"a}schner}}, \bibinfo {author} {\bibfnamefont {B.~S.}\ \bibnamefont
  {Rem}}, \bibinfo {author} {\bibfnamefont {A.}~\bibnamefont {Eckardt}},
  \bibinfo {author} {\bibfnamefont {K.}~\bibnamefont {Sengstock}}, \ and\
  \bibinfo {author} {\bibfnamefont {C.}~\bibnamefont {Weitenberg}},\ }\href
  {https://doi.org/10.1038/s41467-019-09668-y} {\bibfield  {journal} {\bibinfo
  {journal} {Nature Communications}\ }\textbf {\bibinfo {volume} {10}},\
  \bibinfo {pages} {1728} (\bibinfo {year} {2019})}\BibitemShut {NoStop}%
\bibitem [{\citenamefont {Sun}\ \emph {et~al.}(2018)\citenamefont {Sun},
  \citenamefont {Yi}, \citenamefont {Wang}, \citenamefont {Zhang},
  \citenamefont {Sanders}, \citenamefont {Xu}, \citenamefont {Wang},
  \citenamefont {Schmiedmayer}, \citenamefont {Deng}, \citenamefont {Liu},
  \citenamefont {Chen},\ and\ \citenamefont {Pan}}]{PhysRevLett.121.250403}%
  \BibitemOpen
  \bibfield  {author} {\bibinfo {author} {\bibfnamefont {W.}~\bibnamefont
  {Sun}}, \bibinfo {author} {\bibfnamefont {C.-R.}\ \bibnamefont {Yi}},
  \bibinfo {author} {\bibfnamefont {B.-Z.}\ \bibnamefont {Wang}}, \bibinfo
  {author} {\bibfnamefont {W.-W.}\ \bibnamefont {Zhang}}, \bibinfo {author}
  {\bibfnamefont {B.~C.}\ \bibnamefont {Sanders}}, \bibinfo {author}
  {\bibfnamefont {X.-T.}\ \bibnamefont {Xu}}, \bibinfo {author} {\bibfnamefont
  {Z.-Y.}\ \bibnamefont {Wang}}, \bibinfo {author} {\bibfnamefont
  {J.}~\bibnamefont {Schmiedmayer}}, \bibinfo {author} {\bibfnamefont
  {Y.}~\bibnamefont {Deng}}, \bibinfo {author} {\bibfnamefont {X.-J.}\
  \bibnamefont {Liu}}, \bibinfo {author} {\bibfnamefont {S.}~\bibnamefont
  {Chen}}, \ and\ \bibinfo {author} {\bibfnamefont {J.-W.}\ \bibnamefont
  {Pan}},\ }\href {\doibase 10.1103/PhysRevLett.121.250403} {\bibfield
  {journal} {\bibinfo  {journal} {Phys. Rev. Lett.}\ }\textbf {\bibinfo
  {volume} {121}},\ \bibinfo {pages} {250403} (\bibinfo {year}
  {2018})}\BibitemShut {NoStop}%
\bibitem [{\citenamefont {Yi}\ \emph {et~al.}(2019)\citenamefont {Yi},
  \citenamefont {Yu}, \citenamefont {Sun}, \citenamefont {Xu}, \citenamefont
  {Chen},\ and\ \citenamefont {Pan}}]{yi2019observation}%
  \BibitemOpen
  \bibfield  {author} {\bibinfo {author} {\bibfnamefont {C.-R.}\ \bibnamefont
  {Yi}}, \bibinfo {author} {\bibfnamefont {J.-L.}\ \bibnamefont {Yu}}, \bibinfo
  {author} {\bibfnamefont {W.}~\bibnamefont {Sun}}, \bibinfo {author}
  {\bibfnamefont {X.-T.}\ \bibnamefont {Xu}}, \bibinfo {author} {\bibfnamefont
  {S.}~\bibnamefont {Chen}}, \ and\ \bibinfo {author} {\bibfnamefont {J.-W.}\
  \bibnamefont {Pan}},\ }\href {https://arxiv.org/abs/1904.11656} {\bibfield
  {journal} {\bibinfo  {journal} {arXiv:1904.11656}\ } (\bibinfo {year}
  {2019})}\BibitemShut {NoStop}%
\bibitem [{\citenamefont {Satija}\ and\ \citenamefont
  {Zhao}(2016)}]{PhysRevB.94.245128}%
  \BibitemOpen
  \bibfield  {author} {\bibinfo {author} {\bibfnamefont {I.~I.}\ \bibnamefont
  {Satija}}\ and\ \bibinfo {author} {\bibfnamefont {E.}~\bibnamefont {Zhao}},\
  }\href {\doibase 10.1103/PhysRevB.94.245128} {\bibfield  {journal} {\bibinfo
  {journal} {Phys. Rev. B}\ }\textbf {\bibinfo {volume} {94}},\ \bibinfo
  {pages} {245128} (\bibinfo {year} {2016})}\BibitemShut {NoStop}%
\bibitem [{\citenamefont {Lian}\ \emph {et~al.}(2018)\citenamefont {Lian},
  \citenamefont {Sun}, \citenamefont {Vaezi}, \citenamefont {Qi},\ and\
  \citenamefont {Zhang}}]{lian2018topological}%
  \BibitemOpen
  \bibfield  {author} {\bibinfo {author} {\bibfnamefont {B.}~\bibnamefont
  {Lian}}, \bibinfo {author} {\bibfnamefont {X.-Q.}\ \bibnamefont {Sun}},
  \bibinfo {author} {\bibfnamefont {A.}~\bibnamefont {Vaezi}}, \bibinfo
  {author} {\bibfnamefont {X.-L.}\ \bibnamefont {Qi}}, \ and\ \bibinfo {author}
  {\bibfnamefont {S.-C.}\ \bibnamefont {Zhang}},\ }\href {\doibase
  10.1073/pnas.1810003115} {\bibfield  {journal} {\bibinfo  {journal}
  {Proceedings of the National Academy of Sciences}\ }\textbf {\bibinfo
  {volume} {115}},\ \bibinfo {pages} {10938} (\bibinfo {year}
  {2018})}\BibitemShut {NoStop}%
\end{thebibliography}%
\end{document}